\documentclass[12pt]{article}
\usepackage[dvips]{epsfig}
\usepackage[dvips]{graphics}
\usepackage[english]{babel}
\usepackage{amssymb} 
\def\vE{\vec{E}}
\def\VE*{\vec{E}^{*}}
\def\vr{\vec{r}}
\def\vA{\vec{A}}
\def\vx{\vec{x}}
\def\vy{\vec{y}}
\def\nn{\nonumber}
\def\ba{\begin{eqnarray}}
\def\ea{\end{eqnarray}}
\begin{document}
\title{{Remarks on Dirac-like Monopoles, Maxwell and
Maxwell-Chern-Simons Electrodynamics in D=(2+1)}
\author{Winder A. Moura-Melo$^{\mbox{a,}}$
\thanks{Email: winder@cbpf.br.} \hspace{.1cm}and J.A.
Helay\"el-Neto$^{\mbox{a,b,}}$\thanks{Email: helayel@cbpf.br.}
\\ \\$^{\mbox{a}}$\hspace{.1cm}Centro Brasileiro de Pesquisas
F\'{\i}sicas \\ Rua Xavier Sigaud 150 - Urca\\
22290-180, Rio de Janeiro, RJ, Brasil.\\ \\$^{\mbox{b}}$
\hspace{.1cm}Grupo de F\'{\i}sica Te\'orica\\Universidade
Cat\'olica de Petr\'opolis (GFT-UCP)\\Petr\'opolis, RJ, Brasil.}}
\date{} \maketitle
\begin{abstract}
Classical Maxwell and Maxwell-Chern-Simons (MCS) Electrodynamics
in (2+1)D are studied in some details.\ General expressions for
the potential and fields are obtained for both models, and some
particular cases are explicitly solved.\ Conceptual and technical
difficulties arise, however,
for accelerated charges.\ The propagation of electromagnetic signals
is also studied and their reverberation is worked out and discussed.\
Furthermore, we show that a Dirac-like monopole yields a (static)
tangential electric field.\ We also discuss some classical and quantum
consequences of the field created by such a monopole when acting
upon an usual electric charge. In particular, we show that at large
distances, the dynamics of one single charged particle under the
action of such a potential and a constant (external) magnetic field
as well, reduces to that of one central harmonic oscillator, presenting,
however, an interesting angular sector which admits energy-eigenvalues.
Among other peculiarities, both sectors, the radial and the angular one,
present non-vanishing energy-eigenvalues for their lowest levels.
Moreover, those associated to the angle are shown to respond to discrete
shifts of such a variable. We also raise the question on the possibility
of the formation of bound states in this system.\\ \\      
\end{abstract}
\newpage
\section*{Introduction}
Field-theoretic models defined in a (2+1)-dimensional space-time have
been studied for nearly two decades\cite{Sig,DJTAnnals}.\ Actually,
lower-dimensional models have provided many interesting results which
do not take place in the (3+1)D world, e.g., Schwinger' mechanism in
QED$_2$\cite{Schwinger} and fractional statistics in three dimensions
\cite{Fracspin}.\ Consequently, lower-dimensional theories cannot be
considered as mere lower limits of four-dimensional ones; they have
rather revealed characteristics that are intrinsic to its
dimensionality.\\ \\
On the other hand, some (2+1)D theories, whenever supplemented by a
Chern-Simons' term, turn out to exhibit a new interesting physical
content, as for example, Maxwell and Einstein-Hilbert actions
\cite{DJTAnnals,3dtorsion}.\ Furthermore, it has been claimed that such
models (mainly those in the context of MCS) have relevance for a deeper
understanding of some Condensed Matter phenomena, like the Quantum Hall
Effect (QHE)\cite{QHE} and High-Tc Superconductivity \cite{HTCSUP} (see
also, Ref. \cite{tauQED3, paircond}).\\ \\
Although Maxwell and Maxwell-Chern-Simons (mainly the latter, in both
Abelian and non-Abelian frameworks) have attracted a great deal of
efforts, it is curious that one has not provided an ``electrodynamical
body'' (Li\'enard-Wiechert-type potentials, Larmor-like formula and
so forth)  for
such (say, Abelian) theories which would be similar to the one we have for
(3+1)D Maxwell\footnote{Although in a different approach, a classical
analysis of the non-Abelian case (SU(2), more precisely) was performed by
D'Hoker and Vinet\cite{DVinet}.}.\ Thus, we shall try to draw
the attention to the fact that the ``lack'' of a complete
``electrodynamical body'' is related to some serious difficulties, for
instance, in calculating $A_\mu$ (and $F_{\mu\nu}$) for a single
accelerated point-like charge.\ In view of that, a Larmor-like
expression relating energy-flux (radiation) and the acceleration of
the sources is still missing.\\ \\
We start the present work by studying the Maxwell (massless) case.\ Some
results are discussed and a number of difficulties are pointed out.\
Following, we add a Chern-Simons term to the former model and some
consequences of such a procedure are worked out.\ Going on, we analyse the
issue concerning the introduction of a Dirac-like monopole within both
models and some properties of its field.\ Some effects of its potential
on an usual electric charge are discussed in both classical and quantum
(non-relativistic) frameworks.\ We close this paper by pointing out some
Conclusions and Prospects.\\ \\
{\bf i) Classical Maxwell Electrodynamics in D=(2+1)}\\ \\
Let us consider the D=(2+1) Maxwell Electrodynamics (MED$_3$) Lagrangian:
\footnote{Our conventions read: $diag(\eta_{\mu\nu})=(+,-,-)$,
greek letters
running 0,1,2; the 2-D spatial coordinates are labeled by latin letters
running 1,2; and $\epsilon_{012}=\epsilon^{012}=\epsilon_{12}=
\epsilon^{12}=+1$.}
\ba
{\cal L}_{MED} = -\frac14 F_{\mu\nu}F^{\mu\nu}  +j_\mu A^\mu \, .
\label{Lced}
\ea
The invariance of the action under local Abelian gauge transformations,
$A_\mu (x) \rightarrow A_\mu (x) -\partial_\mu \Lambda (x)$, is ensured by
the conservation of the 3-current, say, $\partial_\mu j^\mu =0$.\ Moreover
with the  usual definition of the field strength, $F_{\mu\nu}=\partial_\mu
A_\nu-\partial_\nu A_\mu$, we get $F_{\mu\nu}=(F_{0i}=+(\vE)_i;F_{12}=B)$.\
Next, the field-strength clearly satisfies $\partial_\mu F^{\mu\nu}=j^\nu$
and $\partial_\mu\tilde{F}^\mu=0$, whence there follow:
$$\nabla B=\partial_t\vE^*+\vec{j}^*\,,\quad \nabla\cdot\vE=\rho\quad
{\mbox{\rm and}}\quad \nabla\cdot\vE^*=\partial_t B\,,$$
where we have defined $\tilde{F}^\mu=\frac12\epsilon^{\mu\nu\kappa}
F_{\nu\kappa}=(+B;-\vE^*)$, with the components of a dual-vector given
by $(\vec{U}^*)_i=\epsilon_{ij}U_j$.\\
The dynamical equation for the more basic quantity, $A_\mu$, reads
(in the gauge $\partial_\mu A^\mu=0$): $\partial^2 A_\mu (x)= j_\mu (x)$.
The solutions to this wave-equation may be readily obtained by means of
the well-known Green's function method (or by applying the Hadamard's
{\em Descent Method}, see Ref. \cite{CH} for further details). Such a
function, $G^{2+1}(x-y)$, may be explicitly worked out and reads (the
advanced function is easily got by introducing a $\Theta(-\tau)$;
$\Theta$ is the usual step-function): 
\ba
G^{2+1}_{ret}(x-y)=-\frac{\Theta(\tau)}{2\pi}\int_0^\infty J_0(kr)
\sin(k\tau)dk= -\frac{\Theta(\tau)}{2\pi}\frac{\Theta(\tau^2-r^2)}
{\sqrt{\tau^2-r^2}}\,,\label{G2+1}
\ea
where $\tau=x^0-y^0$ and $r=|\vx-\vec{y}|$.\ The integral above may be
found, for example, in Ref.\cite{Grad} (on page 731 and eq. 6.671-7).
It is worth to notice that $G^{2+1}$ presents a quite different
behaviour respect to its (3+1)D-counterpart, $G^{3+1}$: the support
of $G^{2+1}$ lies no longer on the surface of the light-cone, where
$(x-y)^2=0$, as is the case for $G^{3+1}(x-y)=\delta[(x^0-y^0)^2-
|\vx-\vy|^2]/2\pi$. Indeed, it rather spreads throughout the whole
internal region of the light-cone, where $(x-y)^2>0$ (blowing up as
$(x-y)^2\to0_+$ and vanishing for space-like intervals, $(x-y)^2<0$).
Thus, the Huyghens principle is satisfied by $G^{3+1}$ and violated
by $G^{2+1}$.\\ \\
As we shall see, this will lead to profound
modifications in planar electrodynamics with respect to the (3+1)D
Maxwell theory. For example, by virtue of the failure of Huyghens
principle, electromagnetic signals reverberate in (2+1) dimensions,
and a Larmor-like formula for the radiated power appears to be a
highly non-trivial task.\\ \\
Next, by taking a single point-like charge, $j^\mu (y)= q \int^{+\infty}_
{-\infty} \dot{z}^\mu (s) \delta^{2+1} (y - z(s)) ds$, we get the general
form for its potential (we have omitted the homogeneous part of the
potential):
\ba
A^\mu _{ret}(x)= +\frac{q}{2\pi} \int_{-\infty}^{+\infty} \Theta (x^0 -z^0
(s)) \frac{\Theta [(x -z(s))^2]}{\sqrt{(x-z(s))^{2} }}\, \dot{z}^\mu (s)\,
ds\, , \label{Asol}
\ea
with $(x-z)^2=[(x^0-z^0)^2-|\vx-\vec{z}|^2]$.\ The expression for the
field-strength is also obtained in the usual way, and reads (with
$P=(x-z)^\alpha\dot{z}_\alpha$ and $Q=(x-z)^\alpha\ddot{z}_\alpha$):
\ba
F_{\mu\nu}(x)=\frac{q}{2\pi}\int^{+\infty}_{-\infty}\frac{\Theta(x^0-z^0)
\Theta((x-z)^2)}{P^2\,\sqrt{(x-z)^2}}\left[\ddot{z}_\nu\,(x-z)_\mu\, P
+\dot{z}_\nu\,(x-z)_\mu (1-Q)- \mu\leftrightarrow\nu\right]ds.\label{Fsol}
\ea
Here, it is worthy noticing that, in general, we do not get to solve the
expressions above.\ Actually, we have tried to solve elementary accelerated
motions, say parabolic and hyperbolic ones.\ Unfortunately, we have found
serious difficulties in performing some integrals that are highly
non-trivial and plagued with serious divergences that have to be suitable
handled\footnote{It  was already
pointed out in the literature that (2+1)D Electrodynamics indeed imposes
additional troubles in calculating some quantities; for example,
in Ref.\cite{Ghoshi}, the author discusses some difficulties brought about
by the
logarithmic behaviour of the potential.}.\ In (3+1)D, the scenario is
quite different, because we have a $\delta^{3+1} ((x-z)^2)$ (instead of
$\Theta((x-z)^2)/\sqrt{(x-z)^2}$) which, in turn,
implies in a straightforward factorisation of the integral in $s$-variable,
by picking up only those points for which $(x-z)^2=0$.\\  \\
Hence, we conclude that the ``lack'' of closed analytic expressions for
$A_\mu$ (and $F_{\mu\nu}$) in the case of an arbitrary motion
(Li\'enard-Wiechert-type expressions) is deeply related to
the failure of the Huyghens' principle, since the solutions to the
$\partial^2$-operator in (2+1)D, $G^{2+1}$, do not satisfy such a principle
(indeed, the same happens for any $G^{n+1}$, $n$ even.\ See, for example,
Ref.\cite{CH,CH2,Baker,GeGB}).\\ \\
On the other hand, even the static case (the constant motion may be easily
got by a Lorentz' boost) reveals some of the new characteristics of the
model.\ Thus, by taking $z^\mu=(s,\vec{0})\,\Rightarrow\,\dot{z}^\mu=
(1,\vec{0}),$ we get:
\ba
& A^{\mu}(x)=&\left\{ \begin{array}{l} A^{0}(\vr,t)=-\frac{q}{2\pi}\ln|\vr|
+\frac{q}{2\pi} \lim_{\tau\to +\infty}\left( \ln|\tau +\sqrt{\tau^2 - r^2}|
\right) \\ \vA(\vr,t)=0 \end{array} \right. \label{Aest}\\
&F_{\mu\nu}(x)=& \left\{ \begin{array}{l}F_{0i}(\vr,t)= +\frac{q}{2\pi}
\frac{r^{i}}{r^{2}} -\frac{q}{2\pi}r^{i}\lim_{\tau\to r^{+}}\left(
\frac{\tau}{r^{2}\sqrt{\tau^{2}-r^{2}}}\right)\\ F_{ij}(\vr,t)=0 \end{array}
\right. \label{Fest}\, .
\ea
Here, we notice that, besides the well-known $\ln|\vx|$-behaviour of the
potential in planar Electrodynamics, there is an extra term which explicitly
diverges.\ Such a term clearly represents the asymptotic value of the
potential as $|\vx|\to+\infty$ and is directly related
to the {\em infrared divergence} of the theory.\ Indeed, by calculating
$A_0(x)$ by means of $\tilde{A}_0(k)$ (its Fourier transform),
we may clearly see that such a
term arises when the mass term is set to zero, as below:
\ba
A^{0}(\vr,t)=+\frac{q}{2\pi}\int_{0}^{\infty} \frac{J_{0}(pr)}{p}dp=
\lim_{\mu\to 0_{+}}K_0(\mu{r})\approx-\ln|\vr|-
\lim_{\mu\to 0_{+}}\ln(\mu/2) \, . \nn
\ea\
Whence, we see that as $m\to 0_+$, the last term above blows up. On the
other hand, the
explicitly divergent term appearing in the $F_{\mu\nu}$ above may be
removed by a suitable subtraction procedure, which is possible because
such a quantity vanishes asymptotically.\ [Among others, such subtleties
shall be more explicitly discussed in Ref.\cite{WAJA}].\\ \\
It is interesting to pay attention to the appearance of such an infrared
divergence at the classical level; indeed, infrared problems in (2+1)D
are much more severe than in 4 dimensions. For example, the non-Abelian
case, even in the presence of massive matter, makes sense
only for very special gauge choices \cite{Das}.\\ \\
Still concerning the general $F_{\mu\nu}$-form, eq. (\ref{Fsol}), there
remains
an interesting issue to be pointed out.\ By taking into account the terms
proportional to the acceleration, $\ddot{z}(s)$, which are those that
effectively contribute to the energy-flux and so, to a Larmor-like formula,
we notice that such terms are proportional to $\int ds/\sqrt{(x-z)^2}$,
and might surprisingly lead us to the result
that radiation in (2+1)D no longer falls off with $r^{-1}$. Indeed it may 
increase proportionally to $\ln|\vec{r}|$, as long as $z(s)$ depends on
$s^2$, which is the case for constant accelerated motions.\\ \\
Next, let us point out a rather peculiar characteristic of the model as
long as the propagation of electromagnetic signals is concerned. Let
us start by considering the charge configuration:
$\rho(\vec{y},t')=q\delta^2(\vec{y})\delta(t')$. Its potential reads:
\ba
\Phi_{pulse}(\vx,t)=\frac{q}{2\pi}\frac{\Theta(t-|\vx|)}{\sqrt{t^2-|\vx|^2}}
\,,\label{Phipulse}
\ea
in contrast with its (3+1)D-counterpart $\Phi_{pulse}(\vx,t)=
-q\delta(t-|\vx|)/4\pi|\vx|$. Clearly, although such a signal has been
sharply sent (at $t=0$ it was
just at $|\vx|=0$) it cannot later be recorded as a sharp one: the pulse
develops a ``tail'' (its spreading in time) and so it reverberates.\
Therefore, we now need a {\em very long time to record} a sharp signal
sent at an
earlier time.\ Next, we obtain the superposed case, which is got from
$\rho(\vec{y},t')=q\delta^2(\vec{y})\Theta(t'),$ and reads:
\ba
\Phi_{sup}(\vx,t)=+\frac{q}{2\pi}\ln\left(\frac{t+\sqrt{t^2-|\vx|^2}}
{|\vx|}\right) \Theta(t-|\vx|)\,,\label{Phisup}
\ea
whence we see that these signals superpose in a logarithmic way,
differently from the (3+1) dimensions, where such a superposition
takes place linearly, $\Phi_{sup}(\vx,t)=-q
\Theta(t-|\vx|)/4\pi|\vx|$.\\ \\
The logarithmic superposition leads us to an 
interesting point if we compare with previous results when $t$ is equal or
slightly greater than $|\vx|$: while the single pulse'
potential, eq. (\ref{Phipulse}), appears to
be very strong, the contrary happens to the superposed case, which is
very weak there. However, as time goes by, things straighten up:
while single pulses fall
off, their superposition appears to broaden the potential.\ [The
expressions for the electric
field are also easily obtained and exhibit similar phenomenon concerning
reverberation, while the superposition is ``better-behaved'' than the
$\Phi$-potential].\ Moreover, notice that as (and only as) $t\to\infty$,
we recover the static potential, eq. (\ref{Aest}).\\ \\
Thus, the results discussed above
bring an additional
complication to the (classical, at least) electrodynamics of a system of
interacting charges, since
even single pulses emitted by an electric
charge will demand a very long time to be completely `{\em felt}' by
another one.\ In other words, even the static feature of the potentials
and fields {\em will be no longer} determined only
by the static configuration of the charges. It rather demands a
very long time to actually happen, since at finite times the
electromagnetic quantities are time-dependent.\\ \\
Indeed, in (2+1)D, we may regard the classical propagation of a signal
as if the wave-front travels with velocity $c$, and decreasing in a
such a way that the back point of the signal has null-velocity (this is
exactly what eq. (\ref{Phipulse}) says).\\ \\
Actually, similar conclusions concerning the reverberation of signals were
already discussed by other authors
\cite{CH,Baker}.\ For instance, Courant and Hilbert in their classical
book\cite{CH2} analyse such a propagation and, by virtue of the failure of
the Huyghens principle, they conclude that D'Alembertian' waves (in general),
even if sharply produced, cannot be later recorded with the same
sharpness.\\ \\
Furthermore, we would like here to raise a question in view of what we
have understood about the spreading that unavoidably affects the
classical propagation of sharp
signals in (2+1)D.\ By facing an electromagnetic signal rather as a wave,
reverberation affects its propagation and we can no longer speak of sharp
pulses; on the other hand, if we are to give the electromagnetic signal the
status of a particle, we wonder whether the concept of photon as a localised
energy packet should not be reassessed in the framework of planar
Electromagnetism.\footnote{An analogous question is pertinent in the MCS-case
(next section).\ There, however, by virtue of the mass gap, reverberation
is more expected to happen, since massive (Klein-Gordon or Proca-like) fields
exhibit such a phenomenum even in (3+1) dimensions \cite{Bert,Barut}.\ (See also,
\cite{CFJ} in which is studied a modification of the standard electromagnetism,
by the inclusion of a Lorentz- and Parity-violating Chern-Simons-like term
in (3+1) dimensions).}\\ \\
{\bf ii) Maxwell-Chern-Simons model}\\ \\
Let us write the Lagrangian for the Maxwell-Chern-Simons Electrodynamics
(MCS):
\ba
{\cal L}_{MCS}=-\frac14 F_{\mu\nu}F^{\mu\nu} +\frac{m}{2}
\epsilon^{\mu\nu\kappa}A_\mu\partial_\nu A_\kappa +j_\mu A^\mu
\label{LCS} \, ,
\ea
where $\frac{m}{2}
\epsilon^{\mu\nu\kappa}A_\mu\partial_\nu A_\kappa=\frac{m}{2}
A_\mu\tilde{F}^\mu$ is the (Abelian) Chern-Simons term, which provides a
mass for the boson, $A_\mu$, without breaking the original local gauge
symmetry of the action \cite{DJTAnnals}, $S_{MCS}=\int d^{2+1}x\,
{\cal L}_{MCS}(x)$. Moreover, the mass parameter, $m$, may be taken
to be positive or
negative.\ Depending on the choice of its signal, the `massive photon'
will carry polarisation equal to +1 ($m>0$) or -1 ($m<0$).\footnote{
Talking about {\em spin} in (2+1) dimensions, we should be careful,
since its meaning is rather different from its (3+1)D-counterpart.\
In fact, for a massive particle, its ``spin'' in (2+1)D has
some similarities with the helicity of its massless correspondent
in (3+1)D: only the positive, +1, or negative, -1, polarisations
may take place, while no component of zero-polarisation appears.
See Refs. \cite{BeDJ,Ply1}.} Notice, however,
that in both cases, massless or massive, the ``photon'' carries
{\em only one physical degree-of-freedom}, which highlights its `scalar
nature'.\ Actually, since its mass is given by means of a topological
mass term, we do not expect to have any additional
degree-of-freedom.\\ \\
In a similar way to the massless case, $A_\mu$-potential can be
worked out and reads as below:
\ba
A^\mu(x)=\int d^{2+1}y\left[G^{mass}(x-y)\eta^{\mu\nu} +\frac{m}{m^2}
\left(G^{mass}(x-y)-G^{2+1}(x-y)\right)\epsilon^{\mu\nu\kappa}
\partial_\kappa\right]j_\nu(y)\label{AMCS}\,,
\ea 
where the massive Green' function is given by:$$G^{mass}_{\stackrel{ret}
{adv}}(x-y)=- \frac{1}{2\pi} \frac{\Theta[t^2 -r^2]\cos\left(m\sqrt{t^2
-r^2}\right)}{\sqrt{t^2 - r^2 }}\Theta[\pm\, t].$$with $t=x^0 -y^0$ and
$r=|\vx-\vy|$.\ We clearly see that, as
$m\to 0$, then $G^{mass}\to G^{2+1}$.\ Similarly to its massless
counterpart, $G^{mass}$ does not satisfy the Huyghens' principle:
again, the support spreads throughout the whole region $(x-y)^2\geq 0$.\\
Next, the general expression for $A_\mu$, as produced by a single
point-like charge, takes the form:
\ba
&A^\mu(x)=&+\frac{q}{2\pi} -\int^{+\infty}_{-\infty} \,ds\,\Theta(x^0-z^0(s))
\Theta[(x-z)^2]\left\{\frac{\cos(m\sqrt{(x-z)^2})}
{\sqrt{(x-z)^2}}\dot{z}^\mu +\right.\nn\\
& & \left.+\frac{m}{m^2}\epsilon^{\mu\nu\kappa}\left[\dot{z}_\nu (x-z)_\kappa
\left(\frac{m\sin(m\sqrt{(x-z)^2})}{(\sqrt{(x-z)^2})^2}
+\frac{\cos(m\sqrt{(x-z)^2})-1}{(\sqrt{(x-z)^2})^3}\right)+\right.
\right.\nn\\
& & \left.\left. +\ddot{z}_\nu\dot{z}_\kappa\left
(\frac{\cos(m\sqrt{(x-z)^2})-1}
{\sqrt{(x-z)^2}}\right) \right]\right\}\,  \label{ACS}\,,
\ea
from which we may notice the difficulties which arise in trying to solve
it for arbitrary motions of the charge (indeed, the general solution to
such an expression deeply depends on the massless one).\ There is also a
new sort of term, not present in the massless case, which is explicitly
acceleration-dependent (a radiation-like term, the last one in the eq.
above). Such a term, in turn,
will lead to another one that explicitly
depends on $d^3z/ds^3$ in the expression for $F_{\mu\nu}$: a
back-reaction-like term.\ By virtue of its length, we shall not give the
explicit form for this field here.\ We refer the reader to
Ref.\cite{WAJA}, where a detailed derivation of the results above will
be presented.\ We only anticipate that the possibility that the
radiation increases like a $\ln|\vec{r}|$ also takes place here.\\ \\
Even though a general solution for $A_\mu$ (and $F_{\mu\nu}$) for
arbitrary motions appears to be far off our possibilities, it is
instructive to work out static
quantities which already exhibit some of the new properties brought about
by the Chern-Simons term.\ They read as follows:
\ba
& & A^\mu(x)=\left\{\begin{array}{l}\Phi(\vx)=+\frac{q}{2\pi}K_0(m|\vx|)
\\A^i(\vx)=-\frac{q}{2\pi} \frac{m}{m^2}\frac{\epsilon^{ij}x^j}{|\vx|}
\left(\frac{1}{|\vx|} -m\, K_1(m|\vx|)\right)\end{array}\right.
\label{AMCSest}\,,\\ 
& & F_{\mu\nu}(x)=\left\{\begin{array}{l}E^i(\vx)=-\frac{q}{2\pi}
\frac{mx^i}{|\vx|}K_1(m|\vx|)\\ B(\vx)=+\frac{q}{2\pi}mK_0(m|\vx|)=
m\Phi(\vx)\end{array}\right.\,.\label{FMCSest}
\ea
Now, we see that $A_\mu$ acquires a better asymptotic behaviour:
$A_\mu\to0$ as $|\vx|\to\infty$ (at large distances, $K_0$ and $K_1$
roughly behave as $e^{-|m\vx|}/{\sqrt{|m\vx|}}$).\ Indeed, even the
long-range
sector of $\vec{A}$ now decreases as $|\vx|^{-1}$. Such a sector is
related to the well-known non-dynamical massless pole and also to the
possibility of topological objects such as vortex-like magnetic field.
In addition, due to the Chern-Simons term, the charge now produces a
non-vanishing static magnetic field.\ Nevertheless, this does not lead to
radiation at all.\ Indeed, it is easily to show that $\nabla\cdot
\vec{S}^*=\nabla\cdot(\vec{E}^* B)=0$, with $\vec{S}^*$ being the
Poynting vector.\\ \\
We should now comment on the short-distance behaviour of
these quantities.\ By recalling that, for $|z|\ll 1$, the modified
Bessel functions behave as $K_0(z)\approx -\ln(z/2)$ and $K_1(z)\approx
z^{-1}$, we see that, near the charge, $\Phi$ and $B$ diverge as $\ln|m\vx|$
while $\vec{E}$ blows up as $|\vx|^{-1}$.\ The vector potential, on the other
hand, exhibits a very peculiar behaviour: it vanishes as $|\vx|\to0$! Such a
result is actually in accordance with eq. (\ref{AMCS}): the $A^i$ components
should vanish as $\sqrt{t^2-|\vx|^2}\to 0$.\\ \\
 Moreover, the fact that, as $|\vec{x}|\to 0$, $B\approx \ln|m\vx|$ implies
that a charge within the Chern-Simons framework is a richer object than
within a pure Maxwell context:
along with its massive electric field, it also produces a flux ``tube'' of
magnetic field, of width $m^{-1}$ and strength $q/m$ (what demands $m$ to be
suficiently large). It is precisely in this non-vanishing character of $B$
that there lies the possibility of the fractional statistics exhibited by
such `charges' \cite{WilPis}.\\ \\
Furthermore, it is easy to conclude, using eq. (\ref{AMCS}) for example,
that upon $m\to-m$, $A^0=\Phi$ and $\vec{E}$ remains unchanged while
$\vec{A}$ and $B$ changes their signals.\\ \\
Next, we shall treat the propagation of signals in the Maxwell-Chern-Simons
framework.\ We shall start by obtaining and analysing the single pulse case,
which is produced by $\rho(\vec{y},t')=q\delta^2(\vec{y})\delta(t')$.\ The
quantities read (we have omitted $\Theta(t-|\vx|)$ in all expressions below):
\ba
& & \Phi_{pulse}(\vx,t)=+\frac{q}{2\pi}\frac{\cos(m\sqrt{t^2-|\vx|^2})}
{\sqrt{t^2-|\vx|^2}}\,, \label{PhiMCSsingle}\\ \nn\\ 
& & A_{pulse}^i(\vx,t)=-\frac{q}{2\pi}\frac{m}{m^2}\epsilon^{ij}\partial_j
\left(\frac{\cos(m\sqrt{t^2-|\vx|^2})-1}{\sqrt{t^2-|\vx|^2}}\right)\,, \nn
\ea
for the potentials, while the fields are:
\ba
 & & E_{pulse}^i(\vx,t) =+\frac{q}{2\pi}\partial^i\left
(\frac{\cos(m\sqrt{t^2-|\vx|^2})}
{\sqrt{t^2-|\vx|^2}} \right)+\frac{q}{2\pi}\frac{m}{m^2}\epsilon^{ij}
\partial_t\partial_j
\left(\frac{\cos(m\sqrt{t^2-|\vx|^2})-1}{\sqrt{t^2-|\vx|^2}}\right) \nn\\ 
 & & B_{pulse}(\vx,t)= -\frac{q}{2\pi}\frac{m}{m^2}\nabla^{2}_x
\left(\frac{\cos(m\sqrt{t^2-|\vx|^2})-1}{\sqrt{t^2-|\vx|^2}}\right) \,.\nn
\ea
The reverberation of the pulse is evident: it is very strong when
$t$ is equal or slightly greater than $|\vx|$ and decreases as time
goes by, vanishing as $t\to\infty$.\
The superposed case is obtained by integrating expressions above from $|\vx|$
to $t$.\ For example, the scalar potential superposes as:
\ba
& & \Phi_{sup}(\vx,t)=\int^t _{|\vx|}\Phi(\vx, \tau)d\tau=+\frac{q}{2\pi}
\int^t _{|\vx|}\frac{\cos(m\sqrt{\tau^2-|\vx|^2})}{\sqrt{\tau^2-|\vx|^2}}
d\tau\,, \label{PhiMCSsup}
\ea
Here, a new result takes place in the MCS framework: we cannot exactly
evaluate how electromagnetic signals superpose for arbitrary cases (say,
finite times), since the integral above is not available, in closed form,
unless $t\to\infty$ (the other electromagnetic quantities also depend
on the same integral).\ At this limit, we get (see, for example, Ref.
\cite{Grad}, page 419, eq. 3.754-2):$$\lim_{t\to\infty} \int^t
_{|\vx|}\frac{\cos(m\sqrt{\tau^2-|\vx|^2})}{\sqrt{\tau^2-|\vx|^2}}
d\tau=K_0(m|\vx|),$$ which, in turn, leads us to the static potential, eq.
(\ref{AMCSest}), as $t\to\infty$.\ A similar scenario holds for the other
quantities.\ Thus,
we see that, in the case of the $\vec{E}$-field, only
its longitudinal component survives asymptotically.\\ \\
\begin{sloppypar}
{\bf iii)Dirac-like monopole and its tangential electric field}\\
\end{sloppypar}
Now, let us draw the attention to the introduction of a Dirac-like object
into the previously studied models and to discuss some characteristics and
consequences of the fields produced by this sort of monopole.\\
As it is well-known, such an (point-like) object shows up by breaking the
Bianchi' identity\cite{Dirac}:\footnote{In the Maxwell-Chern-Simons case, the
na\"ive breaking of such an identity yields the breaking of gauge invariance.\
Thus, one should take into account that the monopole induces an extra electric
current in order to balance $\partial_\mu j^\mu=0$, and so restores gauge
invariance (see Ref.\cite{HT,Pisarski} for details.\ See also Ref.\cite{teseetc}
for an alternative approach to a similar problem in (3+1) dimensions).}
$\partial_\mu\tilde{F}^\mu=g$, which in terms of the potentials gets the form:
\ba
\int_{t}dt\int_{xy}d^2 x\,\left(\epsilon_{ij}\,[\partial_i,\partial_t]
A_j(\vx,t) -[\partial_x,\partial_y]\Phi(\vx,t)\right)=g \,;\label{mono}
\ea
in the static limit, it reduces to:
\ba
[\partial_x, \partial_y]\Phi(\vx)=-g\delta^2(\vx) \,.\label{monoest}
\ea
Now, the above equation may be satisfied only if $\Phi$ carries a
`singular structure'.\ Indeed, by recalling that$$ [\partial_x,\partial_y]
\arctan\left(\frac{y}{x}\right)=\partial_x \left(\frac{x}{x^2+y^2}\right)+
\partial_y\left(\frac{y}{x^2+y^2}\right)
$$exactly coincides with$$\nabla^2\ln\sqrt{x^2+y^2}=+2\pi\delta(x)\delta(y),$$
we identically solve eq. (\ref{monoest}) by taking
(as usual $r=\sqrt{x^2+y^2}$ and $\varphi=\arctan(y/x)$)
\ba
\Phi(\vx)=-\frac{g}{2\pi}\arctan\left(\frac{y}{x}
\right)\,\Rightarrow\, \Phi(r,\varphi)=-\frac{g}{2\pi}\varphi \, ,
\label{Phiangular}
\ea
The appearance of the angle-function above suggests us the need for
a single-valuedness requirement: $\Phi(\varphi)=\Phi(\varphi+2\pi n)$.
Its remarkable angular (instead of being radial) dependence leads
to a very interesting (static) electric field ($\vec{\cal E}
=-(\nabla\Phi+\partial_t\vec{A})$, as usual):
\ba
\vec{\cal E}(x,y)=+\frac{g}{2\pi}\frac{x\hat{y}-y\hat{x}}{x^2+y^2}
\,\Rightarrow
\,\vec{\cal E}(r,\varphi)=+\frac{g}{2\pi}\frac{\hat{e}_\varphi}{r}
\,.\label{Etang}
\ea
Whence, we clearly see the announced property of the $g$-monopole:
it yields a (static) tangential electric field\footnote{Strictly speaking,
such a field does not produce a genuine Newton's force on another charge
(usual or peculiar one), since the force between them does not lie on the
line that links both particles, as may be readily seen.}.\ [As far as we
have seen,
such a peculiarity takes place only in (2+1)D Electrodynamics.\ Furthermore,
we do expect that such a property survives at time-dependent regimes]. Moreover,
it is worth noticing that a point-like magnetic vortex is characterised by a
vector potential identical in structure to the tangential electric field
above\cite{Jvortex}. Thus, we may identify a ``duality'' between both
objects: the vortex is obtained from the monopole (more precisely, from
its ``string'' -see below) by taking the electric field and the charge of
the first to be respectively the vector potential and the magnetic flux
associated to the latter.\\ \\
On the other hand,
it is a well-known fact that in (2+1)D the `worldline' of a monopole is
reduced to a point in (2+1) dimensional space-time (see, for example,
Ref.\cite{HT,Pisarski};
see also Ref. \cite{ES}).
Therefore, the singular point above cannot be identified with the monopole
itself. Actually, the
modified Bianchi equation, $\partial_\mu\tilde{F}^\mu=g\delta^2(\vx)$, have
to be rather viewed as an equation for the ``string''  of to the
monopole. What happens is that, at static limit the ``string'' (indeed,
reduced to a spatial point in the (2+1)D-case) appears to be localised
at the origin.\\ \\
Although such a localisation seems to state us that $g$ should be rather
faced as a peculiar electric charge, we stress that this is not so.\ Indeed,
what occurs is that, at static limit, the vanishing of radiation,
$\int\nabla\cdot\vec{S}^*\,d^2 x=\int\nabla\cdot(\vec{\cal E}^*{\cal B})\,
d^2 x=0$, demands that
the monopole' magnetic field must also vanish.\ [Notice that such a
requiriment, ${\cal B}=0$, is intimately related to the tangential
feature of $\vec{\cal E}$, once that $\vec{\cal E}^*$ becomes radial, and so
$\nabla\cdot\vec{\cal E}^*\neq 0$].\ Hence, what we may
state is that such an object yields only non-vashing (tangential) electric
field at the static limit. \\ \\
Next, we analyse the (classical) dynamics of a usual electric charge, $q$,
with mass $m$, moving under the action of such a tangential field.\ Its
equations of motion are easily obtained and read as follows:
\ba
\frac{2\pi m}{gq}\ddot{x}=-\frac{y}{x^2+y^2}\quad {\mbox{\rm
and}}\quad \frac{2\pi m}{g q}\ddot{y}=+\frac{x}{x^2+y^2}\,,
\label{Eqxy}
\ea
or in $(r,\varphi)$-coordinates:
\ba
\frac{2\pi m}{g q}(\ddot{r}-r\dot{\varphi}^2)=0\quad
{\mbox{\rm and}}\quad \frac{2\pi m}{g q}\frac{d}{dt}(r^2\dot{\varphi})=
1\,.\label{Eqrphi}
\ea
Now, due to the angle-dependent feature of the potential, we notice that the
particle' `angular momentum' is clearly not conserved. As far as we have seen,
such a non-conservation imposes an intricate coupling between the coordinates,
what implies in serious difficults towards analytical resolution of the
differential equations.\ A typical plot of the
motion ($x-y$-coordinates) of the charged particle is shown in Fig. 1.\
By virtue of the tangentially repulsive nature of the electric field,
the particle is quickly drifted away, despite the signals of the charges.\\ \\ 
A further system which deserves more attention is that in which
we also have the presence of an external (constant, for concreteness)
magnetic field.\ A realistic planar system may be obtained at very low
temperatures (around or less than 1K) and suficiently strong magnetic
field (at least 10T) perpendicular to a very thin plate
\footnote{Such systems may be realised, for instance, in the interface
between two semi-conductors.\ Furthermore, since the motion of the charges
(electrons, for concreteness) takes place as if the third dimension
(perpendicular to the plane of motion) were frosen, the generally
employed 2D (spatial) treatment is justified, and has been shown to
gives us a very good explanation of the physical phenomena which occur
whithin such systems, e.g., the Quantum Hall Effect.}.\ Such a
perpendicular field is got by taking a vector potential
entirely confined to the 2D-spatial plane, for example,
$\vec{A}=\vec{A_1}=B_0 x\hat{j}\,,\quad \vec{A}=\vec{A_2}=-B_0 y\hat{i}$
(Landau gauges) or still $\vec{A}=\frac12 \vec{A_1} +\vec{A_2}$
(symmetric gauge).\\ \\
Now, our present system is composed by the electric charge subject to
the external magnetic and to the tangential electric field as well.\ Again,
the classical eqs. of motion are easy to be obtained and read (eqs. of motion
in $r,\varphi$ imediately follow):
\ba
\frac{ m}{q}\ddot{x}=-\frac{q}{2\pi}\frac{y}{x^2+y^2}+B_0\dot{y}\quad
{\mbox{\rm and}}\quad \frac{m}{q}\ddot{y}=+\frac{g}{2\pi}\frac{x}
{x^2+y^2}-B_0\dot{x}\,,
\label{Eqxymag}
\ea
Or, by defining complex dynamical variables as $\eta=x+iy$ and $\eta^*=x-iy$,
we get:$$2m(\ddot{\eta}\eta^*+\eta\ddot{\eta}^*) +iqB_0 (\dot{\eta}\eta^*
-\eta\dot{\eta}^*)=0 \quad {\mbox{\rm and}}\quad 4\pi m(\ddot{\eta}\dot
{\eta}^*+\dot{\eta}\ddot{\eta}^*) +iqg \frac{(\dot{\eta}\eta^*
-\eta\dot{\eta}^*)}{\eta\eta^*}=0.$$Despite their symmetric appearance, the
resolution of the eqs. above is not too easy.\ Indeed, we
expect that they may be
even more difficult to be solved than those in the absence of magnetic
field (previous case).\\ \\
On the other hand, numerical resolution shows us
that the magnetic field tends to compensate the repulsive effect of the
electric one so that the (classical) motion of the particle appears
to drift in a slower way, describing an almost regular spiral-like
pattern (see Fig. 2).\ Notice also that the distance between two
neighbour arms of such a pattern decreases as the radial distance
increases: the particle asymptotically `approaches' to perform a closed
trajectory (in the next section, we shall see that, the quantum dynamics
of the charged particle asymptoticaly, $r\to\infty$, reduces to that of one
central harmonic oscillator).\\ \\
There is, however, at least one important information which may be
analytically obtained: in both cases, $B_0=0$ and $B_0\neq0$, the velocity
of the charged particle is bounded by the angle, as below:
\ba
(\vec{v})^2(t)=\frac{qg}{m\pi}\varphi(t) +(\vec{v}_0)^2\,. \label{v2angulo}
\ea
It is worthy noticing that the number of windings of the
charge around the origin must be taken into account, i.e., the
kinectic energy is determined by the total angle descrided by the
charge. [As a sort of quantum
counterpart, we shall see that as $r\to\infty$ the (angular) energy
eigenvalues have to be shifted as $\varphi\to \varphi +2\pi$
(see next section for details)].\\ \\
{\bf iv) Preliminary analysis of the quantum charge-monopole system}\\ \\
Next, we shall present a preliminary (non-relativistic) quantum analysis
of the system above: one electric charge, $q$, moving 
under the action of the monopole scalar potential 
and of an external constant magnetic field, $B_0$.\ The
Hamiltonian (the pure $gq$-system is readily got by setting $\vec{A}=0$),
$$H=\frac{1}{2m}(\vec{p}-q\vec{A})^2 +qV$$ for this system is
obtained by taking $\vec{A}$ in a particular gauge (Landau or symmetric),
as well as
$V(x,y)=-\frac{g}{2\pi}\arctan(y/x)=-\frac{g}{2\pi}\arg(\vec{r})$.\
[Notice that the potential remains invariant under general
scale transformation, say: $x\rightarrow f(x,y)\,x\quad{\mbox{\rm and}}
\quad y\rightarrow f(x,y)\,y\,,$ but, the same symmetry is not
present in the full Hamiltonian, even for $f(x,y)=a=
{\mbox{\rm constant}}$].\\ \\
For the analysis to be presented here, concerning the
non-conservation of the angular-momentum and some of its consequences, as
well as asymptotic bahaviours of the present system, it will be more
convenient to write the Hamiltonian above in polar coordinates, $r,\varphi$,
and $\vec{A}$ in the symmetric gauge, like below:
\ba
H=\frac{1}{2m}\left[p{_r}{^2}+ \frac{p_r}{r}+(qB_0)^2 r^2\right]
+\frac{1}{2m}\frac{p{_\varphi}^2}{r^2}+\frac{qB_0}{2m}p_\varphi
-\frac{gq}{2\pi}\varphi \,, \label{Hrfi}
\ea
with $r$ and $\varphi$ defined as before and $\vec{p}=p_r \hat{e}_r
+\frac{p_\varphi}{r}\hat{e}_\varphi$, whence there follows that
$p_r\leftrightarrow-i\hbar \frac{\partial}{\partial r}$ and
$p_\varphi\leftrightarrow-i\hbar \frac{\partial}{\partial \varphi}$.\\
\\
Now, we notice the first remarkable feature of this Hamiltonian: $H$ is
explicitly angle-dependent and so non-invariant under rotations;
conversely, the angular momentum operator,
$J=p_\varphi=-i\hbar\frac{\partial}{\partial \varphi}$, is not conserved,
$[J,H]=+i\hbar gq/2\pi\neq 0$.\\ \\
Although other angle-dependent Hamiltonians have been studied and shown to
be relevant in Physics (see for example \cite{Hangle}), a remarkable
difference between them and the one presented here is that the latter
is not separable.\ Indeed, as far as
we have seen, the system appears to present an intricate coupling between
its degrees-of-freedom, despite of the coordinates chosen. [Perhaps,
some non-standard tranformation could lead us to such a separation, but
could also lead us, on the other hand, to results which were of hard
physical interpretation.\ Such an issue remains to be investigated].\\ \\
It is clear, from the Hamiltonian (\ref{Hrfi}) and also from the
fundamental commutation relations, $[r,\varphi]=[p_r,p_\varphi]=0$ and
$[r,p_r]=[\varphi,p_\varphi]=+i\hbar$, that the non-separability
arises from the non-conservation of the angular momentum, $[J,H]\neq 0$.\
Indeed, as it may be easily checked, such an angular sector would be
separable if
it had the general form $\frac{1}{r^2}(J^2 +aJ+b\varphi)$.\ So, it is the
lack of a $1/r^2$-factor in $J$ and in $\varphi$-terms what prevents us from
a split of variables.\\ \\
On the other hand, by facing $H$ as being non-separable, the analytical 
resolution of the eigenvalue problem, $H|\psi>=E|\psi>$, appears to be of
very hard achievement.\ [Actually, the presence of the terms proportional
to $\varphi$ and $r$ -or powers of $r$- in $H$ prevents us from solving this
eigenvalue problem by means of, for example, hypergeometric functions (see,
for example Ref. \cite{hyperg})].\\ \\
Therefore, a numerical resolution appears to be a more suitable (and direct)
attempt towards solving the problem (results will be communicated as soon
as they were obtained).\ Here, however, we shall deal with some
analytical results at asymptotic limits, even though some of them appear
to be quite qualitative. We shall mainly discuss the limits
$r\to 0$ and $r\to \infty$:\\ \\
i) $r\to 0$: we have seen that near the origin (where the ``string'' is
localised), the charged particle experiences a very strong tangentially
repulsive electric field (see previous section for details).\ Since 
as $r\to 0$ this field blows up, it is expected that $q$ can never reach
the origin, say, its wave-function must vanish there:
$|\psi(r=0,\varphi)>\equiv 0$. Such a requirement may be viewed as the
counterpart of the Dirac-veto in (3+1)D: a single charge moving under
the action of the magnetic monopole field could not cross the string
of its associated vector potential\cite{Dirac}. [In addition, such a
requirement will impose (see $r\to\infty$-limit below) severe
restrictions on the asymptotic wave solutions].\\ \\
Thus, what remains to be determined is how quickly $|\psi>$ vanishes as
$r\to0$.\ Nevertheless, contrary to the $r\to\infty$-limit, in
which the Hamiltonian gets separable (see below), here the variables
are not
na\"ively separated.\ This arises because $p^2_\varphi/r^2$ is
one of the leading terms, similarly to the original problem, described
by the Hamiltonian (\ref{Hrfi}).\ [In this sense numerical techniques
could help us in order to get some information
about the $gq$-system as $r\to0$, say, the form of the wave-functions
and eigenvalues].\\ \\
ii)$r\to\infty$: supposing that the canonical momenta
remain finite in this limit, we get:
\ba
H(r,\varphi)_{r\to\infty}\approx \frac{1}{2m}(p^2_r +q^2B^2_0
r^2)+\frac{qB_0}{2m}p_\varphi -\frac{gq}{2\pi}\varphi\,,\label{Hrinfty}
\ea
in which the variables appear explicitly split, say,
$H_{r\to\infty}=H^r_{r\to\infty}+H^\varphi_{r\to\infty}$.\ Thus, at this
limit, we have that (the limit $r\to\infty$ is implicit hereafter)
\ba
(H|\psi(r,\varphi)>)=(E_n|\psi(r,\varphi)>)\quad
\Longrightarrow\quad (H^r
+H^\varphi)|R(r)\Phi(\varphi))>=((E^r
+E^\varphi)|R\Phi>)\,,\label{eigenrfi}
\ea
which leads us to:
\ba
H^r R_k(r)=E^r_kR_k(r) \quad{\mbox{\rm and}}\quad
H^\varphi\Phi_l(\varphi)=E^\varphi_l\Phi_l(\varphi) \,.\label{HrHfi}
\ea
Therefore, as $r\to\infty$, we get the following set of differential eqs.:
\ba
& & \hbar^2\frac{d^2}{dr^2}R +(2m E^r -q^2B^2_0r^2)R=0 \,,\label{rlarge}\\
& & i\hbar\frac{d}{d\varphi}\Phi +(\epsilon^\varphi
+\beta\varphi)\Phi=0\,,\label{rlargefi}
\ea
with $\beta=+mg/\pi B_0$ and $\epsilon^\varphi=2mE^\varphi/qB_0$.\\ \\
We notice that, at this limit, the radial part of the Hamiltonian reduces
to that of one central harmonic oscillator, whose solutions may be
written in terms of Hermite polynomials, 
$H_n$:$$R_k(u)=R_0 e^{-u^2/2}H_k(u),$$with $u=qB_0$. This implies the
well-known eigenvalues $E^r_{k}=\hbar\omega_c(k+1/2)$, where
$\omega_c=qB_0/m$ is the cyclotron frequency. Now, by virtue
of the requirement $|\psi(r=0,\varphi)>\equiv 0$, only the
$k={\mbox{\rm odd}}$ solutions will survive. This implies in a
non-vanishing value for the lowest (renormalised) energy
level, $E^1_k=\hbar\omega_c$.\\ \\
On the other hand, the
angular sector appears to be quite unusual.\ Indeed, by solving the
differential equation in $\varphi$, we readily obtain
\ba
\Phi_l(\varphi)=\Phi_0 \,exp\left[\frac{i}{\hbar}\left(\frac{\beta\varphi}{2}
+\epsilon^\varphi_l\right)\varphi\right] \,. \label{Phidefi}
\ea
It is worth noticing the new $\varphi^2$-like phase factor, along with the usual
linear one. As a first remark, we should stress that it cannot be removed by
any suitable gauge tranformation; indeed, it must rather be faced as a
consequence of the $\varphi$-like scalar potential.\ Although quite unusual,
it leads us to new and interesting results. First, notice that $\Phi(\varphi)$
has periodicity $2\pi(\beta\pi
+\epsilon^\varphi_l)$. Thus, the requirement that $\Phi$ be
single-valued, i.e., continuous, is equivalent to set
\ba
2\pi(\beta\pi +\epsilon^\varphi_l)=2\pi l\hbar\quad \Longrightarrow\quad
E^\varphi _l=E^\varphi_0 +\frac{\omega_c}{2}l\hbar\,.\label{angleeing}
\ea
Now, if we identify the parameter $l$ as the number of windings $q$ gives
around $g$ (for example, in the counter-clock-wise sense), then $l$ shall
be taken as a non-negative integer (indeed, the negative values would be
associated to the clock-wise sense). Therefore, the eigenvalues
associated to the angular variable feel whether it is running between
$0$ and  $2\pi$, $2\pi$ and $4\pi$, and so forth. In other words, whenever
$\varphi$ is shifted, say, by $2\pi$, its associated eigenvalues respond to
this change by shifting up their values.\\ \\
However, before completing a winding around $g$, $q$-particle would have
vanishing energy, since $l=0$. It is precisely here that $E^\varphi_0=-gq/2$
enters: since $E^\varphi_0$ is a classical value, the lowest (angular) energy
level of $q$-charge is non-vanishing. In other words, since the angular
potential is acting on $q$, it is expected that, by conservation of energy,
this charge has non-vanishing (also angular) kinetic energy, as long as it
has started its motion. This is the reason why there appears an $E^\varphi_0$
with an intrinsically classical nature.\\ \\
The results above may be viewed as a quantum analogue of eq. (\ref{v2angulo}).
For example, the fact that $\vec{v}^2$ be given by the full angle (including
many windings) is now represented by parameter $l$; and the initial kinetic
energy, $\vec{v}^2_0$ (which may be classically set to zero) `survives' at
quantum level, but acquiring a intrinsic non-vanishing value.\\ \\
Moreover, we could be tempted to naively apply $J$-operator on $|\psi_{kl}>$
above, to get$$J|\psi_{kl}>=(\beta\varphi
+\epsilon^\varphi_l)|\psi_{kl}>=(\beta(\varphi+\pi)+l)|\psi_{kl}>,$$
and hence, to guess that $|\psi_{kl}>$ carry continuous angular momentum.
However, this is not a legitimate procedure, because $|\psi_{kl}>$ are
not eigenvectors of $J$ (recall that $[J,H]\neq0$).\ Actually, as far
as we have seen, the only two quantities which may be simoutaneously
diagonalised in $|\psi_{kl}>$-basis are $H_r$ and $H_\varphi$ (the
components of the asymptotic Hamiltonian, eq. (\ref{Hrinfty})).\\ \\
Clearly, the results and remarks above are strictly valid only at the
asymptotic limits specified previously.\ Whether similar scenario does
happen at
arbitrary distances (as the classical result (\ref{v2angulo})
does), remains to be studied and will be
strongly dependent on the separation of variables in the original
Hamiltonian, eq. (\ref{Hrfi}).\\
A naive analysis of the limits discussed above would lead us to conclude
that, since the charged particle is
repelled from the origin by the $\varphi$-potential and since as
$r\to\infty$ its dynamics reduces to that of one central
harmonic oscillator (whose wave-functions fall off exponentially), it is
expected that the system yields physical bound states.\ Therefore, even
though the pure $gq$-system does not admit bound states (once that the
confining $r^2$-type potential is absent, for this case, in
eq.(\ref{Hrinfty}), we get indeed a radially free particle), when it is
supplemented by a suitable external magnetic field, the possibility for
such states may be raised.\\ \\
Nevertheless, when electrons are moving on the
plane subject only to a perpendicular magnetic field, then the choice of
Landau gauge immediately reduces the quantum problem to that of one
harmonic oscillator in one dimension, and a free particle motion
in the other direction. In this case, we cannot have bound
states.\footnote{Nevertheless, whenever two species of fermions
are combined into a unique four-component spinor, the presence of
a constant magnetic field induces flavor symmetry breakdown and
fermion condensates appear\cite{KeGMS}. Such condensates are,
however, quite sensible to thermal effects and disappear at
finite temperature \cite{DasHott}.} However, when
the system is supplemented by an extra, say, scalar potential (as in
the present case), it is also well-known that bound states show up,
even in the case of repulsive potential\cite{MorPran}.\ Here, we have
just raised such a question, and a precise answer demands further
investigation.\\ \\   
{\bf v) Conclusion and Prospects}\\ \\
We have shown that classical (2+1)D Maxwell and Maxwell-Chern-Simons
Electrodynamics present some interesting novelties as compared to
Maxwell theory in (3+1)D, namely, the reverberation of signals and the
far-from-trivial question of a Larmor-like formula. As we have seen,
such phenomena are intimately related to the failure of the Huyghens'
principle.\ Namely, the latter is very difficult to be obtained even
for constant accelerated motions (parabolic and hyperbolic ones).\ The
integrals involved are highly
non-trivial and appear to diverge, so demanding some suitable
regularisation scheme.\ On the other hand, we hope that some hints about
such a Larmor' formula could be obtained with the help of numerical
calculations. Next, as a natural extension of our present results,
we shall pursue an
investigation of the canonical quantisation of the electromagnetic
radiation for the models contemplated here \cite{winprog}.\\ \\
Concerning the Dirac-like monopole, it also presents some new properties
whenever compared to its (3+1)D-counterpart; for instance, its static
tangential electric field.\ Furthermore, acting on a single
charged particle, it leads us to interesting classical and quantum
results. For example, the $gq$-system (with $B_0$) has been shown to
give rise, at least asymptoticaly and at non-relativistic regimes, to
a central harmonic oscillator, with an interesting angular sector which
contributes to the energy-eigenvalues.\\ \\
As future prospects, solutions to the Hamiltonian of eq. (\ref{Hrfi})
in its general form shall be the object of a further investigation
\cite{HJAWA}. It woul be also of relevance to compute possible effects
of this peculiar potential on spin particles, for instance,
planar Dirac fermions. Moreover, by virtue of its peculiar scalar
potential (and unusual
consequences), such a monopole could be relevant to Condensed Matter
problems. For instance, by looking at this object as a sort of impurity
(scatter) within a sample, could its presence modify the Hall
conductivity? And eventually, how would such a
modification actually look like?\\ \\
{\centerline{\bf  Acknowledgements}}\\ \\The authors are grateful to
Prof. S.A. Dias, Prof. B. Schroer, F. Araruna, H. Belich, J.L. Boldo,
R. Casana, G. Cuba Castillo, O. Del Cima, R. Klippert, L. Moraes,
A. Nogueira, R. Paunov  and R. Rodrigues for useful discussions.\
They also express their gratitude to Prof. M. Henneaux for a careful
reading of an earlier version of this manuscript and for having drawn their
attention to the work of Reference\cite{DVinet}. Prof. Ashok Das and Prof. M.
Plyushchay are deeply ackowledged for a number of very pertinent comments and
for having drawn our attention to some relevant references. Finally, the
authors would also like to thank Prof. R. Jackiw for having pointed out the
work of Reference\cite{CFJ}. CNPq-Brasil is also acknowledged for the
financial support.\\

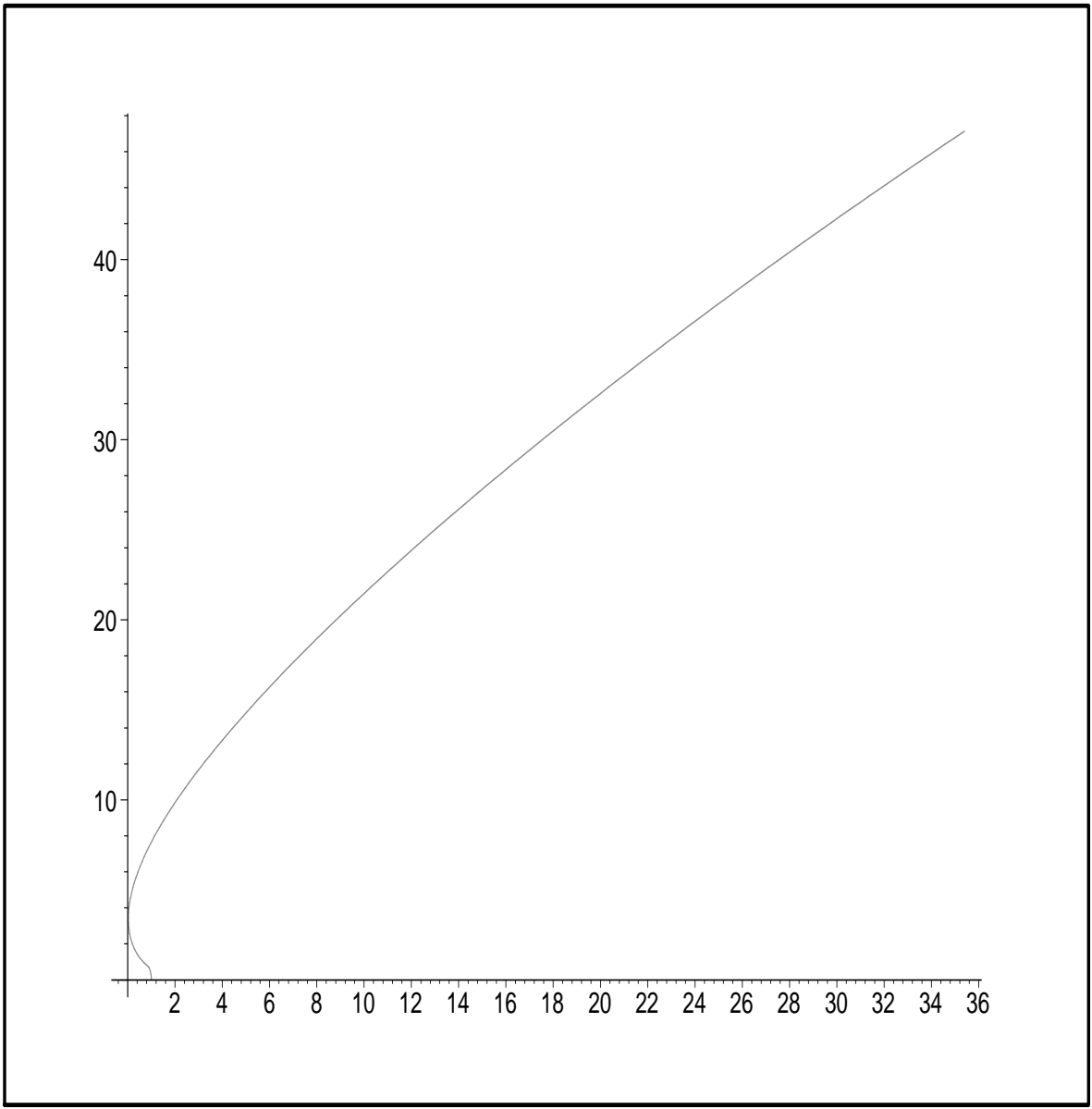
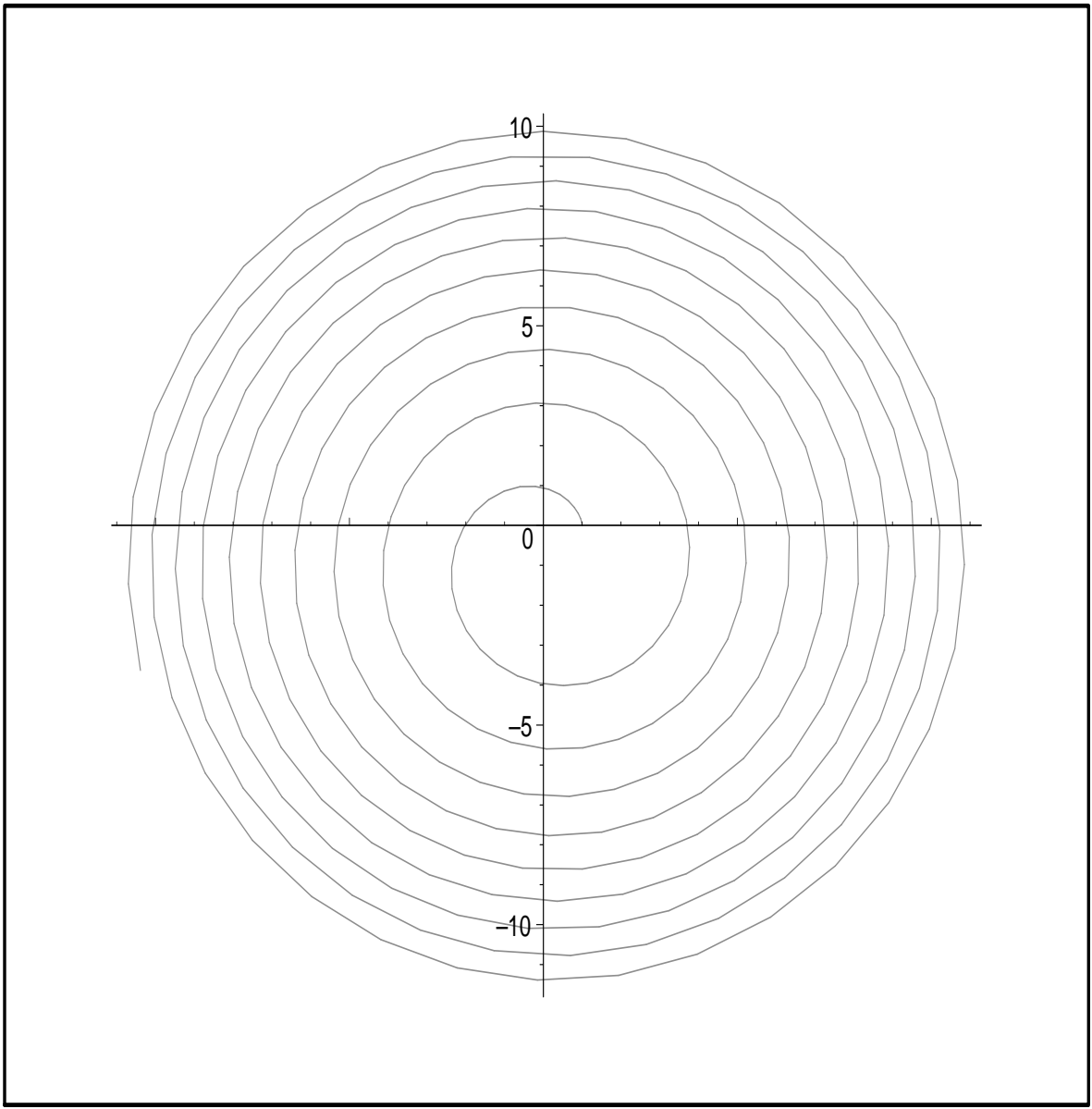
\end{document}